\documentclass[preprint,12pt]{elsarticle}
\usepackage[utf8]{inputenc}
\usepackage[english]{babel}
 
\usepackage[dvipsnames]{xcolor}

\usepackage{mathenv}

\usepackage{axodraw}

\usepackage{color}

\definecolor{myhigh}{HTML}{228b22}
\definecolor{myhighB}{HTML}{6a5acd}

\usepackage{amssymb}

\journal{Physics Letters B}

\begin{document}
\def\spa#1.#2{\left\langle#1\,#2\right\rangle}
\def\spb#1.#2{\left[#1\,#2\right]}

\def\tree{\rm tree}

\def\la{\langle}
\def\ra{\rangle}

\newcommand{\oneloop}{{\rm 1-loop}}
\def\Mloop{M^{\oneloop}}
\def\Mtree{M^{\tree}}
\def\Atree{A^{\tree}}
\def\Aloop{A^{\oneloop}}

\def\Mcount{M^{\rm c}}

\def\dLIPS{\int d LIPS\; }
\def\eps{\epsilon}

\def\notag{\nonumber}

\def\II{{\cal I}}
\def\GG{{\cal G}}
\def\FF{{\cal F}}
\begin{frontmatter}



\title{Loop Amplitudes in an Extended Gravity Theory}


\author{David~C.~Dunbar, John~H.~Godwin, Guy~R.~Jehu and Warren~B.~Perkins}

\address{College of Science, \\
Swansea University, \\
Swansea, SA2 8PP, UK}

\begin{abstract}
We extend the $S$-matrix of gravity by the addition of the minimal three-point amplitude or equivalently adding $R^3$ terms to the Lagrangian. 
We demonstrate how Unitarity can be used to simply examine the renormalisability of this theory and determine the $R^4$ counter-terms that arise at one-loop.  
We find that the combination of $R^4$ terms that arise in the extended theory is complementary to the $R^4$ counter-term  associated with supersymmetric 
Lagrangians. 
\end{abstract}


\end{frontmatter}


\section{Introduction}

There are different approaches to defining a quantum field theory.  One approach is to specify a local Lagrangian density containing fundamental fields
and then use the associated Feynman diagrams to determine the scattering  amplitudes that contribute to the $S$-matrix.
Alternatively, one can 
use the on-shell amplitudes to define the theory~\cite{tHooft:1973wag}.  In this approach singular structure and symmetries are crucial to constrain  and 
ultimately compute the scattering  amplitudes. With enough constraints, higher point amplitudes may be derived from a limited set of lower point 
functions~\cite{Eden}. 
In the minimal case all $n$-point amplitudes may be constructed from only the three-point amplitudes plus the requirement of factorisation. In 
\cite{Benincasa:2007xk} such theories were described as {\it fully constructible}.   Yang-Mills and gravity both have fully constructible
tree amplitudes. In \cite{Dunbar:2017dgp} we showed that in a theory of gravity deformed by additional three point vertices the leading deformation
to the tree amplitudes was also fully constructible.

In this article we examine the one-loop amplitudes for both extended Yang-Mills and gravity
and in  particular their ultra-violet structure. 
In agreement with long established results we find that the extended Yang-Mills theory 
is renormalised~\cite{Narison:1983kn, Morozov:1985ef, Bauer:1997gs, Gracey:2002he}.
For gravity,  we find that additional 
counter-terms/amplitudes are required.  
The use of  four-dimensional unitarity techniques  gives relatively easy access to the ultra-violet structures of these theories.

\section{Structure of the Amplitudes} 

A  one-loop
amplitude in a theory of massless particles can be expressed, after a
Passarino--Veltman reduction~\cite{PassVelt}, in the form
\begin{equation}
\Aloop_n=\sum_{i\in \cal C}\, a_i\, I_4^{i} +\sum_{j\in \cal D}\,
b_{j}\, I_3^{j} +\sum_{k\in \cal E}\, c_{k} \, I_2^{k} +R_n +O(\epsilon) 
\label{DecompBasis}
\end{equation}
where the $I_m^i$ are $m$-point scalar integral functions and the $a_i$
etc. are rational coefficients. 
${\cal C}$ is the set of box integral functions with all allowed partitions of the external legs between the corners. 
For a color-ordered Yang-Mills amplitude the allowed partitions respect the cyclic ordering of the legs.
Similarly ${\cal D}$ and ${\cal E}$
are the sets of triangle and bubble integral functions. 
$R_n$ is a purely rational term.  The integral functions depend upon the number of ``massive'' legs 
(or more correctly legs with non-null momenta) so, for example, the triangle functions come in three types : with either one, two or three massive legs.
This form is an expansion in terms of the integral functions appearing and so is useful when computing the 
coefficients from the cut singularities of the amplitude.

Alternately, we can re-express the one-loop amplitude for pure Yang-Mills or gravity
in a form which highlights the singular structure of the amplitude,
\begin{equation}
\Aloop_n= \Atree_n  \II_n +\GG_n+\FF_n  +{\cal R}_n
\end{equation}
where  $\II_n$ contains the soft-singular Infra-Red (IR)~\cite{Kunszt:1994np} terms of the amplitude and is
\begin{equation}
\II_n = -\frac{c_\Gamma}{(4\pi)^2}\sum_{i=1}^n  \frac{(-s_{ii+1}/\mu^2)^{-\eps}}{\eps^2}
\end{equation}
for a colour ordered gluon amplitude where  $c_\Gamma=(4\pi)^{\epsilon}\Gamma^2(1-\epsilon)\Gamma(1+\epsilon)/\Gamma(1-2\epsilon)$. 
For a graviton scattering amplitude~\cite{Weinberg:1965nx,Dunbar:1995ed}, 
\begin{equation}
\II_n = -\frac{c_\Gamma}{(4\pi)^2}\sum_{i<j} s_{ij} \frac{(-s_{ij}/\mu^2)^{-\eps}}{\eps^2}
\; . 
\end{equation}

Within the decomposition of (\ref{DecompBasis})
the contributions to $\II_n$ arise from the box integral functions and the one and two mass triangle integral functions.   $\GG_n$ is of the form
\begin{equation}
\GG_n =   \sum_i  c_i { (-s_{ii+1}/\mu^2 )^{-\epsilon}\over \eps} 
\end{equation}
for gluon scattering amplitudes and 
\begin{equation}
\GG_n =  \sum_{i<j} c_{ij} { (-s_{ij}/\mu^2 )^{-\epsilon} \over \eps} 
\end{equation}
for graviton scattering.
Within the integral basis decomposition~(\ref{DecompBasis}), the $\GG_n$ arise from the bubble integral functions.  The $\GG_n$ terms contain both the collinear IR singular terms and the UV divergences. 
The function $\FF_n$ contains the finite transcendental functions. These arise from both the box integral functions and from the three-mass triangle integral 
function.   ${\cal R}_n$ is the remaining finite rational term. 

The coefficients of the integral functions can be determined using four dimensional generalised unitarity techniques from the on-shell amplitudes. 
Computing ${\cal R}_n$ from unitarity requires using $d=4-2\epsilon$ tree amplitudes.

We will find that the form of the IR singularities is not altered in the extended theories: as might be expected by naive power counting. 

\section{Yang-Mills Case}

Before looking at gravity theories, we consider the case of gluon scattering in pure Yang-Mills.
We will consider a color ordered formalism and examine the color ordered partial amplitudes which have cyclic symmetry. 
The full amplitude can be reconstructed by multiplying by the color factors and summing over permutations.  
We also restrict to the  leading in $N_c$ color ordered partial amplitude for which the one-loop leading in color amplitudes have a factor of $N_c$.

We  define the theory from the fundamental three point amplitudes, 
\begin{equation}
A( 1^{h_1} , 2^{h_2} , 3^{h_3} )
\end{equation}
where $h_i$ is the helicity of the $i$-th leg. For gluons the helicity is $\pm 1$. 
Three point amplitudes vanish for real momenta but may be non-zero if we allow complex momenta.  If we express
the momenta in spinor variables $p^{\alpha\dot{\alpha}}=\lambda^{\alpha}\bar\lambda^{\dot{\alpha}}$ then 
amplitudes become functions of the 
bilinears $\spa{a}.b =\epsilon_{\alpha\beta} \lambda_a^{\alpha}\lambda_b^{\beta}$ and
$\spb{a}.b =-\epsilon_{\dot\alpha\dot\beta} \bar\lambda_a^{\dot\alpha}\bar\lambda_b^{\dot\beta}$. 
Since, for a three point amplitude of massless particles we must have
$s_{ab}=\spa{a}.b\spb{b}.a=0$,
there are two possibilities:
\begin{eqnarray}
a) \;\;\;  \spa{i}.{j}=0  ,\;\; \spb{i}.{j} \neq 0
\notag 
\\
b) \;\;\; \spb{i}.{j}=0  ,\;\; \spa{i}.{j} \neq 0
\end{eqnarray}
Consequently we can build three point amplitudes either from $\spa{i}.j$ or $\spb{i}.j$ but not both.  
Under scaling 
$\lambda_i \longrightarrow t_i \lambda_i$,
$\bar\lambda_i \longrightarrow t_i^{-1} \bar\lambda_i$ the amplitude must scale as $t^{-2h_i}$.
Finally requiring that the amplitude vanishes for real momenta leads to the following unique non-zero  three-point gluon amplitudes,
\begin{eqnarray}
A_{3:{\rm tree}}(1^-,2^-,3^+) &=& g{ \spa1.2^3 \over \spa2.3\spa3.1}\;,
\nonumber\\
A_{3:{\rm tree}}(1^+,2^+,3^-) &=& g{\spb2.1^3 \over \spb2.3\spb3.1}\;,
\nonumber\\
A_{3:{\rm tree}}(1^+,2^+,3^+) &=& \alpha g \spb1.2\spb2.3\spb3.1\;,
\nonumber\\
A_{3:{\rm tree}}(1^-,2^-,3^-) &=& \alpha g \spa1.3\spa3.2\spa2.1\;.
\label{vertdef}
\end{eqnarray}
The first two amplitudes are the well known MHV (``Maximally Helicity Violating'') and $\overline{\rm MHV}$ amplitudes. They form the top
elements in a multiplet involving all the helicities of N=4 super-Yang-Mills. The parameter $\alpha$ is dimensionful with mass dimension minus two, ie. $\alpha=c/M^2$ where
$c$ is dimensionless and $M$ is some mass scale\footnote{With the normalisation of (\ref{vertdef}) each $n$-point, $L$-loop amplitude contains a factor of $g^{n-2+2L}$ 
 which we suppress.} and the $\alpha$ expansion can be considered as an expansion in inverse powers of the 
mass $M$.  

In this letter 
we consider  expanding the theory by including the $\alpha$-vertices in addition to the MHV vertices.
The amplitudes in this theory can then be expanded as a power series in $\alpha$,
\begin{equation}
A_n(1,\cdots , n) =  \sum_{r=0}  {\alpha}^r  A_n^{\alpha^r}(1,\cdots , n)
\end{equation}
where $A_n^{\alpha^0}$ is the usual Yang-Mills amplitude.  

From a Lagrangian field theory viewpoint we are extending the theory by 
\begin{equation}
{\cal L}_{F^3}= \alpha' {\rm Tr} ( F_{\mu\nu}F^{\nu\rho} F_{\rho}{}^\mu ) \;.
\end{equation}
where $\alpha'=-\alpha g /3$~\cite{Broedel:2012rc}. 
Having defined the three-point vertices, factorisation can be used~\cite{Cohen:2010mi,Dunbar:2017dgp} 
to obtain the leading four-point tree amplitudes:
\begin{eqnarray}
A^{\alpha}_{4:{\rm tree}}(1^+,2^+,3^+,4^+)&=& 2 \alpha{stu \over \spa1.2\spa2.3\spa3.4\spa4.1}
= 2 \alpha  K_{++++} \times u \;,
\nonumber\\
A^{\alpha}_{4:{\rm tree}}(1^-,2^+,3^+,4^+)&=& -\alpha{ \spb2.4^2 s t \over \spb1.2\spa2.3\spa3.4\spb4.1}
=\alpha   K_{-+++} \times u \;.
\end{eqnarray}
(These expressions match those obtained from Feynman diagram calculations~\cite{Dixon:1993xd}, or 
color-kinematics duality~\cite{Broedel:2012rc} or scattering equations~\cite{He:2016iqi})
The combinations $K_{++++}$ and $K_{-+++}$ carry all the necessary spinor weight of the amplitude with
$|K_{++++}|=|K_{-+++}|=1$ for real momenta.  
The factor $K_{++++}$ has manifest cyclic symmetry but is also fully crossing symmetric since
\begin{equation}
{st \over \spa1.2\spa2.3\spa3.4\spa4.1}={su\over \spa1.2\spa2.4\spa4.3\spa3.1}={tu \over \spa1.4\spa4.2\spa2.3\spa3.1}\;.
\end{equation}
Similarly $K_{-+++}$ has manifest flip-symmetry but is also invariant under exchange of the positive helicity legs, 
$2\leftrightarrow 3$ etc. These factors can be written in many ways, e.g.
\begin{equation}
K_{++++} = -{ \spb1.2^2 \over \spa3.4^2} = -{ \spb1.2^2\spb3.4^2 \over s^2}\;.
\end{equation}

For the pure Yang-Mills theory, all tree amplitudes can be constructed using factorisation from the three-point trees~\cite{Britto:2005fq}: 
i.e. the theory is 
``constructible''  using the definition of \cite{Benincasa:2007xk}.   
The leading deformed amplitudes $A_n^{\alpha}$ also can also be constructed in this way, however amplitudes beyond this leading deformation are not 
constructible purely from factorisation. This will be pursued further in the context of gravity theories later. 

We now wish to determine the one-loop amplitudes in the extended theory to leading order in $\alpha$. 
Unitarity methods have proven very efficient in determining  one-loop amplitudes using the on-shell tree amplitudes.

Working in four dimensions the simplicity of the four-dimensional trees greatly simplifies the calculation of the coefficients of the integral functions,  
but does not allow us to compute the finite rational terms ${\cal R}$.  Since we are mainly interested in the UV singularities, which come with 
an accompanying $\ln(\mu^2/s)$,  four dimensional  two-particle cuts suffice~\cite{Norridge:1996he}.   

Noting that
\begin{eqnarray}
A^{\alpha}_{4:1-{\rm loop}}(1^-,2^-,3^+,4^+)|_{cut\; part} 
&=& 0
\end{eqnarray}
since there are no non-vanishing four dimensional cuts for these amplitudes,  the non-vanishing $\alpha^1$ amplitudes are the all-plus and the single minus.

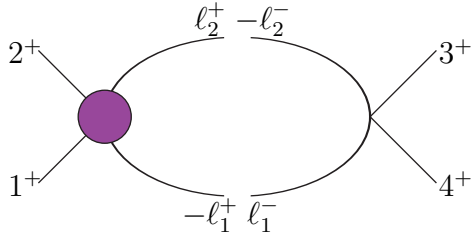
\begin{figure}[h]
\centering
\begin{picture}(150,100)( 0,0)   
\COval(75,50)(30,50)(0){Black}{White} %
\Line(25,50)(0,75)
\Line(25,50)(0,25)
\Line(125,50)(150,75)
\Line(125,50)(150,25)
\COval(25,50)(10,10)(0){Black}{Purple} 
\SetWidth{10}
\SetColor{White}
\Line(75,0)(75,100)
\Text(-5,75)[c]{$2^+$}
\Text(-5,25)[c]{$1^+$}
\Text(158,75)[c]{$3^+$}
\Text(158,25)[c]{$4^+$}
\Text(65,87)[c]{$\ell_2^+$}
\Text(65,13)[c]{$-\ell_1^+$}
\Text(85,87)[c]{$-\ell_2^-$}
\Text(85,13)[c]{$\ell_1^-$}
 \end{picture}  
    \caption{The two-particle cut of the  all-plus amplitude. }
    \label{fig:bibbleallplus}
\end{figure}

Calculating the $s$-channel two-particle cut for the all-plus amplitude, as shown in fig.\ref{fig:bibbleallplus}, gives
\begin{eqnarray}
C_2&=&  \dLIPS  A^{\alpha}_{4:{\rm tree}}(1^+,2^+,\ell_2^+,-\ell_1^+) \times A^{\alpha^0}_{4:{\rm tree}}(\ell_1^-,-\ell_2^-,3^+,4^+)
\nonumber \\
&=&  -2\alpha \dLIPS{ s_{12}s_{1\ell_1}s_{1\ell_2} \over \spa1.2\spa2.{\ell_2}\spa{\ell_2}.{\ell_1} \spa{\ell_1}.{1} }
\times { \spa{\ell_1}.{\ell_2}^4 \over \spa3.4\spa4.{\ell_1}\spa{\ell_1}.{\ell_2} \spa{\ell_2}.{3 } }
\nonumber \\
&=& - 2\alpha \dLIPS { s_{12}  \over  \spa1.2 \spa3.4} { s_{1\ell_2} \spb{\ell_1}.{1} \over\spa2.{\ell_2} } 
\times { \spa{\ell_1}.{\ell_2}^2 \over \spa4.{\ell_1} \spa{\ell_2}.{3 } }
\label{eq:eq16}\end{eqnarray}
where we also have a configuration where the $\alpha$ vertex is on the right hand side. 
Manipulating eq.~(\ref{eq:eq16})  using 
\begin{equation}
{ \spa{\ell_2}.{\ell_1} \spb{\ell_1}.{1} \over \spa2.{\ell_2} }
={ \spa{\ell_2}.{2} \spb{2}.{1} \over \spa2.{\ell_2} } =\spb1.2\;,
\end{equation}
we have
\begin{eqnarray}
C_2&=& - 2 \alpha \dLIPS  {  \spb1.2^2 \over   \spa3.4}
\times {  s_{1\ell_2} \spa{\ell_1}.{\ell_2} \over \spa4.{\ell_1} \spa{\ell_2}.{3 } }\;.
\end{eqnarray}
Now
\begin{equation}
\spa4.{\ell_1} \spa{\ell_2}.{3 } \times \spb{\ell_1}.{\ell_2} = \la 4| \ell_1 \ell_2 | 3 \ra = -\spa4.3 [3|\ell_2|3 \ra = -\spa3.4 
(\ell_2-k_3)^2\;,
\end{equation}
so
\begin{eqnarray}
C_2&=& 2\alpha  \dLIPS {  \spb1.2^2  s_{12} \over   \spa3.4^2 }
\times {  [1|\ell_2|1\ra  \over   (\ell_2-k_3)^2 }\;.
\end{eqnarray}
Now, rather than perform the integration over the cut momenta we recognise this as the cut of a 
covariant integral which we can evaluate:
\begin{equation}
\dLIPS  f(\ell_1,\ell_2) \longrightarrow \int d^D\ell  { f(\ell_1,\ell_2) \over \ell_1^2 \ell_2^2 }
\end{equation}
where we only keep terms with a $s$-channel cut in the resultant integral~\cite{Bern:1994zx,Bern:1994cg}.
Consequently we replace $C_2$ by a triangle integral with linear numerator $I_3[\ell^\mu_2]$.  Using
\begin{equation}
I_3[\ell^\mu_2 ]=  { (k_4-k_3)^\mu \over s} I_2 (s)+    k_3^\mu I_3(s)\;,
\label{eq:lineartriangle}
\end{equation}
where $I_2(s)$ is the scalar bubble  integral function and  $I_3(s)$ is the one-mass 
scalar triangle integral function which only 
depend on the kinematic variable $s$,  
we obtain
\begin{eqnarray}
& &-2\alpha{  \spb1.2^2   \over   \spa3.4^2 }  [1(4-3)|1\ra I_2(s) 
-2\alpha {  \spb1.2^2   s_{12} \over   \spa3.4^2 }  [1|3|1\ra I_3(s)  
\notag \\
&=& -2\alpha{  \spb1.2^2   \over   \spa3.4^2 }   ( t-u) I_2(s) 
-  2\alpha{  \spb1.2^2   s_{12} s_{13} \over   \spa3.4^2 }  I_3(s) 
\notag \\
&=&  2 \alpha K_{++++}  ( t-u) I_2(s) 
+  2 \alpha K_{++++} u \times s  I_3(s) \;.
\end{eqnarray}
Doubling this to account for inserting the $F^3$ operator on the opposite side of the cut 
and combining with the $t$-channel cut leads to
the amplitude
\begin{eqnarray}
A^{\alpha}_{\oneloop} (1^+,2^+,3^+,4^+) & & =  A^{\alpha}_{\tree}  {\cal I}_4
\notag \\
+
4 \alpha K_{++++}   & & 
\Bigl[  (t-u) I_2(s)+(s-u)I_2(t) \Bigr] 
+{\cal R}_4
\end{eqnarray}
where we have used
\begin{equation}
2 s  I_3(s) +2t I_3(t)=  {\cal I}_4\;.
\end{equation}
Note that there are no box functions in this amplitude and the triangle functions generate the soft part of the IR term. 
The overall  coefficient of  $\epsilon^{-1}$ is
\begin{equation}
4{\alpha\over (4\pi)^2}  \left(  (t-u)+(s-u) \right) K_{++++} =-{12\over (4\pi)^2}  u \alpha  K_{++++} 
=-{6 \over (4\pi)^2}  \times A^{\alpha}_{\tree}
\label{eq:bareallplus}
\end{equation}
which is proportional to the tree 
and so the one-loop UV infinity leads to a renormalisation of the $\alpha F^3$ term.

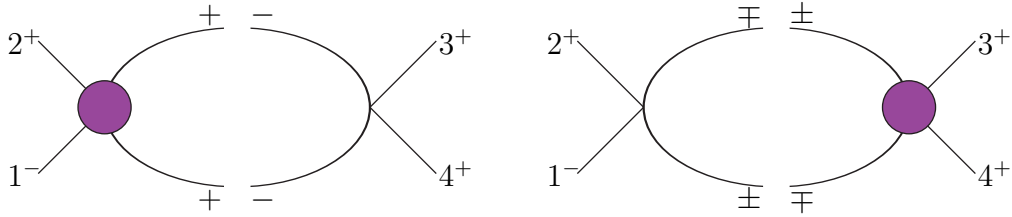
\begin{figure}
  \begin{picture}(200,100)( 0,0)   
\COval(75,50)(30,50)(0){Black}{White} %
\Line(25,50)(0,75)
\Line(25,50)(0,25)
\Line(125,50)(150,75)
\Line(125,50)(150,25)
\COval(25,50)(10,10)(0){Black}{Purple} 
\SetWidth{10}
\SetColor{White}
\Line(75,0)(75,100)
\Text(-5,75)[c]{$2^+$}
\Text(-5,25)[c]{$1^-$}
\Text(158,75)[c]{$3^+$}
\Text(158,25)[c]{$4^+$}
\Text(65,85)[c]{$+$}
\Text(65,15)[c]{$+$}
\Text(85,85)[c]{$-$}
\Text(85,15)[c]{$-$}
 \end{picture}  
 \begin{picture}(150,100)( 0,0)   
\COval(75,50)(30,50)(0){Black}{White} %
\Line(25,50)(0,75)
\Line(25,50)(0,25)
\Line(125,50)(150,75)
\Line(125,50)(150,25)
\COval(125,50)(10,10)(0){Black}{Purple} 
\SetWidth{10}
\SetColor{White}
\Line(75,0)(75,100)
\Text(-5,75)[c]{$2^+$}
\Text(-5,25)[c]{$1^-$}
\Text(158,75)[c]{$3^+$}
\Text(158,25)[c]{$4^+$}
\Text(65,85)[c]{$\mp$}
\Text(65,15)[c]{$\pm$}
\Text(85,85)[c]{$\pm$}
\Text(85,15)[c]{$\mp$}
 \end{picture}  
     \caption{The bubbles for single minus}
    \label{fig:bubbloneminus}
\end{figure}

We also compute the one-loop contribution to the single minus amplitude $A^{(1)}(1^-,2^+,3^+,4^+)$. 
There are three non-zero configurations as shown in fig.~\ref{fig:bubbloneminus}. The first of these is 
\begin{eqnarray}
C_2^A &=&  A_{4:{\rm tree}}^{\alpha} (1^-,2^+,l_2^+,-l_1^+) \times A^{\alpha^0}_{4:{\rm tree}} ( l_1^-,-l_2^-,3^+,4^+)
\notag
\\
&=&  -\alpha { \spb2.{-l_1}^2 s_{12} s_{2l_2} \over \spb1.2\spa2.{l_2}\spa{l_2}.{-l_1} \spb{-l_1}.1 } \times 
{ \spa{l_1}.{-l_2}^3 \over \spa{-l_2}.3\spa3.4\spa{4}.{l_1} }
\notag 
\\
&=& -\alpha { \spa1.2 \over \spa3.4}   { \spb2.{l_1}^2  \spb{2}.{l_2} \over  \spb{l_1}.1 } \times 
{ \spa{l_1}.{l_2}^2 \over \spa{l_2}.3\spa{4}.{l_1} }\;.
\end{eqnarray}
This can be rearranged to 
\begin{equation}
C_2^A  =   - \alpha {  s_{12}^2  \over \spa3.4^2 }   \times { \spb2.{l_1}  \spb{2}.{l_2}  \spa{l_1}.1\spa{l_2}.{1} \over  
(l_1-k_1)^2(l_1+k_4)^2 }
\end{equation}
which is the two particle cut of a quadratic box. Replacing this by a covariant integral and only keeping terms with a $s$-channel cut we obtain 
\begin{equation}
C_2^A= - \alpha K_{-+++} 
\Biggl(  \frac{s^3 t}{2u} I_4 +\frac{s^3}{u} I_3(s) -suI_3(s)-(2t+s)  I_2(s) 
\Biggr)\;.
\end{equation}
The other terms are a little more complex (requiring a quartic box integral) and yield
\begin{eqnarray}
C_2^B &=& -\alpha K_{-+++} 
\Biggl(  \frac{s t^3}{2 u} I_4(s,t) +\frac{st^2}{u} I_3(s)
-t I_2(s) \
\Biggr) \;,
\notag \\
C_2^C  &=& -\alpha K_{-+++} \Biggl( \frac{stu}{2}  I_4(s,t)  +u I_2(s) \Biggr)\;.
\end{eqnarray} 
Combining these gives
\begin{eqnarray}
A^{\alpha}_{\oneloop} (1^-,2^+,3^+,4^+)  =
- \alpha  K_{-+++} \biggl( & &\frac{st}{2u} (s^2+t^2+u^2)I_4(s,t)+\frac{(s^2+t^2-u^2)}{u}(s I_3(s)+tI_3(t)) 
\notag  \\ 
& &- (2s+4t) I_2(s)-(4s+2t) I_2(t) ) \biggr) +{\cal R}_4\;.
\end{eqnarray}
The box functions combine with the triangle functions 
to generate the IR singular terms and the finite transcendental function, 
\begin{equation}
a I_4 + \sum_{s,t}  b_j I_3^j = \Atree_4 {\cal I}_4 +{\cal F}_4
\end{equation}
where
\begin{equation}
\FF_4=  -\frac{\alpha}{(4\pi)^2} K_{-+++} \times {  (s^2+u^2+t^2)  \over  u }  \ln^2(s/t)
\end{equation}
and we find no corrections to the IR structure. 

We also have the UV terms: the coefficient of $\epsilon^{-1}$ in these terms is
\begin{equation}
\frac{\alpha}{(4\pi)^2} K_{-+++} \left( 2s+4t)+(4s+2t)  \right) = -6 \frac{\alpha}{(4\pi)^2}u K_{-+++} 
= -\frac{6}{(4\pi)^2} A_{tree}^{\alpha} (1^-,2^+,3^+,4^+)
\label{eq:baresingleminus}
\end{equation} 
which matches the singularity found for the $A(1^+2^+3^+4^+)$ case.

Equations~(\ref{eq:bareallplus}) and (\ref{eq:baresingleminus}) are the singularities in the 
bare amplitudes.
The amplitudes contain universal collinear IR singularities~\cite{Kunszt:1994np}:
\begin{equation}
-\frac{n\beta_0 g^2}{2(4\pi)^2} A_{tree}^{\alpha}  =-\frac{22 N_c g^2}{3 (4\pi)^2} A_{tree}^{\alpha}
\end{equation}
where $\beta_0=11N_c/3$. We determine the UV divergence by first subtracting these from the bare singularity.

When renormalising the theory there must be a simultaneous renormalisation of $g^2$ and $\alpha$. The renormalisation of $g^2$ is unaltered  
since it is determined by the $\alpha^0$ amplitudes.

\begin{equation}
g^2 \alpha \longrightarrow g^2 \alpha -g^4 \beta_0  \alpha + g^2 \delta\alpha
\end{equation}
so that (reinserting $g$ and $N_c$)
\begin{equation}
g^2 \delta\alpha - \frac{g^4 \beta_0 \alpha}{\epsilon} -\frac{6N_c\alpha}{(4\pi)^2\epsilon}  
=-\frac{ 22 N_c g^2}{3 (4\pi)^2} \alpha
\end{equation}
so that
\begin{equation}
\delta\alpha = \frac{g^2N_c \alpha}{(4\pi)^2\epsilon} ( 6 -\frac{11}{3} ) =\frac{7N_c\alpha }{3}  \frac{g^2}{(4\pi)^2}
\end{equation}
This value of $7N_c/3$ matches previous calculations of the anomalous dimension of Yang-Mills extended 
by the $F^3$ operator~\cite{Narison:1983kn, Morozov:1985ef, Bauer:1997gs, Gracey:2002he}.

\section{Extended Gravity  Amplitudes}

We now consider extending gravity by additional three point vertices. 
For gravitons with $h=\pm 2$ the possible three point amplitudes 
are\footnote{We remove a factor of $i(\kappa/2)^{n-2+2L}(4\pi)^{-(2-\epsilon)L}$ from the $n$-point $L$-loop amplitude.}
\begin{eqnarray}
M_{3:{\rm tree}}(1^-,2^-,3^+) &=& { \spa1.2^6 \over \spa2.3^2\spa3.1^2}\;,
\nonumber\\
M_{3:{\rm tree}}(1^+,2^+,3^-) &=& {\spb1.2^6 \over \spb2.3^2\spb3.1^2}\;,
\nonumber\\
M_{3:{\rm tree}}(1^+,2^+,3^+) &=& \alpha \spb1.2^2\spb2.3^2\spb3.1^2\;,
\nonumber\\
M_{3:{\rm tree}}(1^-,2^-,3^-) &=& \alpha \spa1.2^2\spa2.3^2\spa3.1^2\;.
\end{eqnarray}
From a Lagrangian viewpoint we are considering the theory
\begin{equation}
L= \int d^D x \sqrt{-g} (  R +  {\alpha\over 60 }  R_{abcd}R^{cdef}R_{ef}{}^{ab}   )\;.
 \end{equation}
The non-vanishing four-point  $\alpha^0$ tree is
\begin{equation}
{M}^{\alpha^0}_{4:{\rm tree}}(1^-,2^-,3^+,4^+) =  -{ \spb3.4 \spa1.2^6 \over\spa3.4 \spa1.3 \spa1.4 \spa2.3\spa2.4  }
\end{equation} and the non-zero $\alpha^1$ trees are
\begin{eqnarray} 
{M}^{\alpha}_{4:{\rm tree}}(1^+,2^+,3^+,4^+) &=& -10 
\left( { s t \over \spa1.2\spa2.3\spa3.4\spa4.1 } \right)^2 { stu }
\nonumber
\\
&=& -10  K_{++++}^2 \times stu\;,
\nonumber
\\
{M}^{\alpha}_{4:{\rm tree}}(1^-,2^+,3^+,4^+) &=& 
\left( { \spb2.4^2  \over \spb1.2\spa2.3\spa3.4\spb4.1 } \right)^2 
{ -s^3 t^3 \over   u } 
\nonumber
\\
&=&  -K_{-+++}^2 \times stu \;.
\end{eqnarray}
These four-point amplitudes due to a $R^3$ term have been computed using field theory methods
long ago~\cite{vanNieuwenhuizen:1976vb} but can also be obtained from the three-point amplitudes 
by factorisation~\cite{Cohen:2010mi,Dunbar:2017dgp} .
These expressions also appear as the  UV infinite
 pieces of both  
two-loop gravity in four dimensions~\cite{Dunbar:2017qxb,Bern:2017puu} and one-loop gravity in six dimensions~\cite{Dunbar:2002gu}.

We evaluate the one-loop amplitude for the all-plus amplitude via the two particle cuts
\begin{eqnarray}
C_2 &  = &   \dLIPS M^{\alpha}_{4:{\rm tree}}( 1^+,2^+,\ell_2^+,-\ell_1^+) \times M^{\alpha^0}_{4:{\rm tree}}( \ell_1^-,-\ell_2^-,3^+,4^+)
\end{eqnarray}
Arranging the trees carefully we obtain
\begin{eqnarray}
C_2 &  = &   {10} \dLIPS { \spb1.2^2 \spb{\ell_1}.{\ell_2}  \over \spa{\ell_1}.{\ell_2} } {  \spb2.{\ell_1}\spb{1}.{\ell_2} } {\spb2.{\ell_2}\spb{\ell_1}.1}
\times { s \spa{\ell_1}.{\ell_2}^4   \spb3.4 \over \spa3.4^3 } \left( \frac{1}{s_{\ell_14}}+\frac{1}{s_{\ell_13}}  \right)
\nonumber
\\
&=&10  \dLIPS 
{s^2  \spb1.2^2  \spb3.4\ \over \spa3.4^3   } {  \spb2.{\ell_1}\spb{1}.{\ell_2} } {\spb2.{\ell_2}\spb{\ell_1}.1}
\times { \spa{\ell_1}.{\ell_2}^2   } \left( \frac{1}{s_{\ell_14}}+\frac{1}{s_{\ell_13}}  \right)
\nonumber\\
&=& {10}  \dLIPS 
{s^2  \spb1.2^4  \spb4.3\ \over \spa3.4^3   } {  [1|\ell_1|1\ra [1|\ell_2|1\ra }
\times \left( \frac{1}{s_{\ell_14}}+\frac{1}{s_{\ell_13}}  \right)
\end{eqnarray}
which is the cut of a pair of quadratic triangle integrals. Replacing this by the covariant integral and evaluating yields
\begin{equation}
{10}{ tu \over s^3 }  \spb1.2^4  \spb3.4^4  \times s^2 I_3(s) +{10}{  \spb1.2^4  \spb3.4^4 \over s^2  }  \times \left( {t^2}+{u^2}-4tu  \right) I_2(s)\;,
\end{equation}
where both integrals yield the same resultant integral function.
After 
doubling 
this to account for inserting the $R^3$ operator on the opposite side of the cut,
this can be rewritten
as 
\begin{equation}
20 K_{++++}^2 stu \times  s^2 I_3(s) +20 K_{++++}^2 \times  s^2 \left(  {t^2}+{u^2}-4tu \right) I_2(s)\;.
\end{equation}
The triangle functions correctly generate the IR terms and
so the the amplitude can be expressed as 
\begin{equation}
M^{\alpha}_{4:{\rm 1-loop}}=  M^{\alpha}_{4:{\rm tree}} \II_4 +\FF_4+\GG_4  +{\cal R}_4
\end{equation}
with $\FF_4=0$ and 
\begin{eqnarray}
\GG_4= 20 K_{++++}^2 \times \Biggl[ s^2 \left( {t^2}+{u^2}-4tu \right) & &I_2(s) + t^2 \left( {s^2}+{u^2}-4su  \right) I_2(t)
\notag \\
+& & u^2  \left( {t^2}+{s^2}-4ts \right) I_2(u) \Biggr] \;.
\end{eqnarray}

The coefficient of  $\epsilon^{-1}$ is 
\begin{equation}
20 K_{++++}^2  \times \biggl(  s^2 \left( {t^2}+{u^2}-4tu  \right)+\{s\leftrightarrow t\} +\{s \leftrightarrow u\} \biggr)
={10} K_{++++}^2 \times (s^2+t^2+u^2)^2 \;.
\label{eq:pureUV}\end{equation}
In Einstein gravity the IR singularity is of the form $\sum s\ln(s)/\epsilon$~\cite{Weinberg:1965nx,Dunbar:1995ed}
and the additional vertex will not affect this by power counting. 
Therefore the rational $\epsilon^{-1}$ singularities in eq.~(\ref{eq:pureUV}) represent the ultra-violet divergence. 
Unlike the Yang-Mills case this is not a renormalisation of the cubic vertex but must be cancelled by the 
addition of a four-point vertex produced by a higher-dimension local operator.

As a consistency check  we also consider the single minus amplitude.  
The $s$-channel bubble in this case has three configurations:
\begin{eqnarray}
 M^{\alpha}_{4:{\rm tree}}(1^-,2^+,\ell_2^+,-\ell_1^+) &\times & M^{\alpha^0}_{4:{\rm tree}}( \ell_1^-,-\ell_2^-,3^+,4^+) \;,
\\
 M^{\alpha^0}_{4:{\rm tree}}(1^-,2^+,\ell_2^{\mp},-\ell_1^\pm) & \times & M^{\alpha}_{4:{\rm tree}}( \ell_1^\pm,-\ell_2^\mp,3^+,4^+)\;.
\end{eqnarray}
These give contributions to the coefficients of the bubble integral functions.  We find
\begin{eqnarray}
c_A=K_{-+++}^2 \times {s^2 \over t u } \left( 2(t^4+u^4) +5ut( t^2+u^2) \right)\;,
\notag\\
c_B=K_{-+++}^2 \times {s^2 \over t u }  \left( 2(t^4+u^4) -3ut( t^2+u^2) \right) 
\end{eqnarray}
giving the overall coefficient of $I_2(s)$ to be
\begin{eqnarray}
c=K_{-+++}^2 \times {s^2 \over t u } \left( 4(t^4+u^4) +2ut( t^2+u^2) \right)\;.
\end{eqnarray}
Extracting the UV divergence we find
\begin{eqnarray}
& &\frac{1}{\eps} \times \left({s^2 \over t u } \left( 4(t^4+u^4) +2ut( t^2+u^2) \right)+\{ s \leftrightarrow u \} +\{ s \leftrightarrow t \} 
\right)
\notag \\
 & &=0
\end{eqnarray}
and so this amplitude has no UV divergence.

In summary, the UV infinities for the four-point one-loop amplitudes are, (re-inserting the appropriate factors)
\begin{eqnarray}
M^{\alpha}_{4:{\rm 1-loop}}(1^-,2^-,3^+,4^+)\bigg|_{1/\epsilon} &=& 0 \;,
\notag \\
M^{\alpha}_{4:{\rm 1-loop}}(1^+,2^+,3^+,4^+)\bigg|_{1/\epsilon}  &=&
\alpha\frac{i}{(4\pi)^{2}}
 \left( \frac{\kappa}{2}\right)^{4}
 \frac{1}{\epsilon}  \times {10} K_{++++}^2 \times (s^2+t^2+u^2)^2 \;,
\notag \\
M^{\alpha}_{4:{\rm 1-loop}}(1^-,2^+,3^+,4^+)\bigg|_{1/\epsilon}  &=& 0 \;.
\end{eqnarray}
Unlike the Yang-Mills case, the UV infinity is not removed by a renormalisation of the three-point vertex but requires 
the addition of a four-point vertex 
which acts as a counterterm. In the following section we examine the counterterm that is required.

\section{$R^4$ operators}
From~\cite{Fulling:1992vm}
the general $R^4$ counterterm in arbitrary dimension is
\begin{equation}
\frac{i}{(4 \pi)^4 \eps} \, \bigg[
a_1 T_1 + a_2 T_2 +a_3 T_3 +a_4 T_4 + a_5 T_5 +a_6 T_6 +a_7 T_7
\bigg]
\end{equation}
where
\begin{eqnarray}
T_1 &=& ( R_{abcd} \, R^{abcd})^2 \;,
\cr
T_2 &=& R_{abcd} \, R^{abc}{}_e \, R_{fgh}{}^d \, R^{fghe}\;,
\cr
T_3 &=& R^{ab}{}_{cd} \, R^{cd}{}_{ef} \, R^{ef}{}_{gh} \, R^{gh}{}_{ab}\;,
\cr
T_4 &=& R_{abcd} \, R^{ab}{}_{ef} \, R^{ce}{}_{gh} \, R^{dfgh}\;,
\cr
T_5 &=& R_{abcd} \, R^{ab}{}_{ef} \, R^c{}_g{}^e{}_h \, R^{dgfh}\;,
\cr
T_6 &=& R_{abcd} \, R^a{}_e{}^c{}_f \, R^e{}_g{}^f{}_h \, R^{bgdh}\;,
\cr
T_7 &=& R_{abcd} \, R^a{}_e{}^c{}_f \, R^e{}_g{}^b{}_h \, R^{fgdh}
\end{eqnarray}
and the combination
\begin{equation}
-{ T_1 \over 16}
+{T_2}
-{T_3  \over  8}
-T_4+2T_5-T_6 +2T_7
\end{equation}
vanishes on-shell in any dimension due to it
being proportional to the Euler form
\begin{equation}
\eps_{a_1a_2a_3a_4a_5a_6a_7a_8}
\eps^{b_1b_2b_3b_4b_5b_6b_7b_8}
R^{a_1a_2}{}_{b_1b_2}
R^{a_3a_4}{}_{b_3b_4}
R^{a_5a_6}{}_{b_5b_6}
R^{a_7a_8}{}_{b_7b_8}\;.
\end{equation}

In $D=4$ these $R^4$ tensors
reduce  to two independent tensors. One of these is the square of the
 ``Bel-Robinson'' tensor~\cite{BelRob}
which was shown to be consistent with
supersymmetry and thus became a candidate counterterm for supergravity
theories \cite{R4sugra}. 
In higher dimensions this tensor
extends to a two-parameter set~\cite{DS}. 

Computing with the general counterterm~\cite{Dunbar:2002gu,TurnerThesis} 
\footnote{this is in arbitrary dimension $D$ where the momenta of the four particles define a four dimensional plane and we can choose two of the helicities to lie in this plane. }
\begin{eqnarray}
\Mcount(1^{+},2^{+},3^{+},4^{+}) &=&
8 \, (8 a_1 + 2 a_2 + 4 a_3 + a_6)
\times  { (s^2 + t^2 + u^2 )^2  }
  K^2_{++++} \;,
\notag
\\
\Mcount(1^{-},2^{+},3^{+},4^{+}) &=&
0\;,
\notag
\\
\Mcount(1^{-},2^{-},3^{+},4^{+}) &=& 8(16 a_1 + 4a_2 + 4 a_4+ 3 a_6+2 a_7) \times \spa1.2^8  K_{++++}^2
\end{eqnarray}
where we see explicitly the two independent choices of tensors expressed in amplitudes. 
In general a tensor could be expressed as 
\begin{equation}
 c_+ R^4_+ +c_- R^4_-
\end{equation}
where $R^4_+$ is the counterterm consistent with supersymmetry and $R_-^4$ is {\it orthogonal} to it, in the  sense that it yields 
$\Mcount(1^{-},2^{-},3^{+},4^{+})=0$. A general counterterm would be a combination of the two. 
 
Clearly the order $\alpha^1$ theory is made UV finite by the addition of the orthogonal counterterm, $R_-^4$. 
This could be realised, for example, by choosing $R_-^4=T_3$ with
\begin{equation}
R^4_{counter} =  \frac{i}{(4 \pi)^4 \eps}   T_3 
\end{equation}

\section{Beyond Cubic Vertices}  

\begin{figure}[h]
\includegraphics{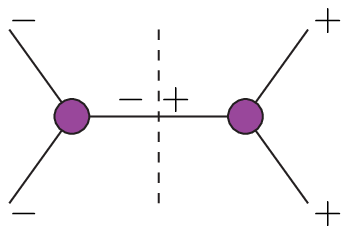}
    \caption{Factorisations of the four-point MHV amplitude at $\alpha^2$.}
    \label{fig:alphasquared}
\end{figure}

The non-extended theory of graviton scattering ( and of gluons) is {\it constructible}: that is the entire $S$-matrix can be generated by 
demanding that the amplitudes are factorisable~\cite{Benincasa:2007qj}. 
In practice the factorisation can be excited by the BCFW shift. 
In the  extended theory  the leading deformation of the $S$-matrix is 
also constructible~\cite{Dunbar:2017dgp}
albeit by using alternative shifts~\cite{Risager:2005vk,BjerrumBohr:2005jr}.
However at order $\alpha^2$, if we consider 
$M_{4:{\rm tree}}^{\alpha^2}(1^-,2^-,3^+,4^+)$ there is a single factorisation as shown in fig.~\ref{fig:alphasquared}.  The amplitude
\begin{equation}
M^{\alpha^2}_{4:{\rm tree}} (1^-,2^-,3^+,4^+) =  \alpha^2 \spa1.2^4\spb3.4^4 \left(  {tu+\beta s^2 \over s } \right)
\end{equation}
has the correct factorisation for any choice of $\beta$.  This ambiguity means we also have to specify the four-point amplitude to 
determine the $S$-matrix. 
In the diagrammar approach this ambiguity
arises due to the existence of a polynomial function with the correct symmetries and spinor and momentum weight.  
From a field theory perspective, additional counterterms can contribute to this amplitude. Specifically,  we could deform the theory via
\begin{equation}
R \longrightarrow R + C_\alpha  R^3  + C_\beta D^2 R^4
\end{equation}
and the four-point amplitude is only specified once $C_{\alpha}$ and $C_{\beta}$ are determined. 

From a constructibility viewpoint, defining the theory from its four-point amplitudes is much less constrained than using the three-point amplitudes because 
momenta constraints do not limit the vertex to be constructed only from $\lambda_i$ or $\bar\lambda_i$ but can involve both (or more likely momenta invariants).  
Specifically we could introduce a fundamental amplitude
\begin{equation}
M_{4:{\rm tree}} (1^-,2^-,3^+,4^+) = \beta' s  \spa1.2^4\spb3.4^4 
\end{equation}
From a Lagrangian perspective this would be implemented by a $D^2R^4$
operator  giving non-vanishing $M(1^-,2^-,3^+,4^+)$ but  vanishing  $M(1^+,2^+,3^+,4^+)$ and $M(1^-,2^+,3^+,4^+)$. As the all-plus and 
single-minus amplitudes vanish in a supersymmetric theory, any operator compatible with supersymmetry that generates a non-vanishing four-point
MHV amplitude will suffice. 
A specific choice of this is~\cite{Dunbar:1999nj}
\begin{equation}
t_{10} t_8 \partial^2 R^4
 =
t_{10}^{a_1a_2a_3a_4a_5a_6a_7a_8a_9a_{10}}
t_8{}_{b_1b_2b_3b_4b_5b_6b_7b_8} 
\partial_{a_1} R_{a_2a_3}{}^{b_1b_2}
 \partial_{a_4} R_{a_5a_6}{}^{b_3b_4}
R_{a_7a_8}{}^{b_5b_6}
R_{a_9a_{10}}{}^{b_7b_8}
\end{equation}
where 
\begin{equation}
t_{10}^{a_1a_2a_3a_4a_5a_6a_7a_8a_9a_{10}}
=  \delta^{ a_1a_4}\delta^{a_2a_5}
\delta^{a_3a_6}\delta^{a_7a_9}\delta^{a_8a_{10}}
-4  \delta^{a_1a_4}\delta^{a_2a_{10}}\delta^{a_3a_5}
\delta^{a_6a_7}\delta^{a_8a_9}
\end{equation}
and, where possible, this should be anti-symmetrised with respect to the 
pairs of indices $a_2\leftrightarrow a_3 $ etc. and symmetrised with respect to
pairs of couples of indices $(a_2a_3) \leftrightarrow (b_1b_2)$~\cite{Dunbar:1999nj}. 

Also, $t_8\,=\,{ 1 \over 2 } \Bigl(t_{(12)}+t_{(48)} \Bigr) $
with 
\def\sym4#1#2#3#4{(\delta^{#1#3}\delta^{#2#4}-\delta^{#1#4}\delta^{#2#3}) }
\begin{eqnarray}
t_{(12)}^{ijklmnpq}=&- 
\biggl( 
(\delta^{ik}\delta^{jl}-\delta^{il}\delta^{jk})
(\delta^{mp}\delta^{nq}-\delta^{mq}\delta^{np})
+\sym4{k}{l}{m}{n}\sym4{p}{q}{i}{j}
\notag  \\
& \null\hskip 2.0 truecm 
+\sym4{i}{j}{m}{n}\sym4{k}{l}{p}{q}
\biggr)
 \\
t_{(48)}^{ijklmnpq}=&
\biggl( \delta^{jk}\delta^{lm}\delta^{np}\delta^{qi}
+\delta^{jm}\delta^{nk}\delta^{lp}\delta^{qi}
+\delta^{jm}\delta^{np}\delta^{qk}\delta^{li}
+[i\leftrightarrow j]+[k\leftrightarrow l]+[m\leftrightarrow n] 
\notag \biggr)
\end{eqnarray}
where $[i\leftrightarrow j]$ denotes antisymmetrisation with respect to
$i$ and $j$. The tensor $t_8$ appears in many situations involving maximal supergravity~\cite{Green:1987mn} and $t_{10}$ appears in loop amplitudes of less than maximal supergravity~\cite{Dunbar:1999nj}.

Although constructability from three-point vertices is an attractive concept, unfortunately we find the theory is not completely specified by the three-point vertex. 

\section{Conclusions}

We have studied the $S$-matrix of extended Yang-Mills and gravity using a diagrammar approach 
in which the theory is defined by its on-shell amplitudes. 
If we wish to extend either pure Yang-Mills or gravity by the addition of a three-point interaction there is an essentially unique
choice.  This choice leads to a theory in which the leading deformation is constructible from three-point amplitudes although 
higher order deformations require further information to fix the amplitudes.   In this letter we have studied  the one-loop
corrections to these theories and demonstrated how Unitarity can be used to simply examine the renormalisability. 
For Yang-Mills the  one-loop UV infinities renormalise the three-point vertex  at leading order.  For gravity however the UV infinities must
be cancelled by four-point amplitudes arising from a different source. 
Extending the $S$-matrix of gravity by the addition of the minimal three-point amplitude is  equivalent to adding $R^3$ terms to the Lagrangian. 
From a Lagrangian view point this is then renormalised at one-loop by $R^4$ counterterms. 
For the leading deformations we find that these counterterms are the combination that is orthogonal to the $R^4$
counterterm associated with supersymmetric Lagrangians.

\section{Acknowledgements}

This work was supported by STFC grant ST/L000369/1.  GRJ was supported by STFC grant ST/M503848/1.
JHG was supported  by the College of Science (CoS)
Doctoral Training Centre (DTC) at Swansea University.

 \appendix


\begin{thebibliography}{00}

\bibitem{tHooft:1973wag}
  G.~'t Hooft and M.~J.~G.~Veltman,
  NATO Sci.\ Ser.\ B {\bf 4} (1974) 177.
  
\bibitem{Eden}
R.J. Eden, P.V. Landshoff, D.I. Olive, J.C. Polkinghorne, {\it
The Analytic S Matrix}, (Cambridge University Press, 1966).


\bibitem{Benincasa:2007xk}
  P.~Benincasa and F.~Cachazo,
  arXiv:0705.4305 [hep-th].
  
  
\bibitem{Dunbar:2017dgp}
  D.~C.~Dunbar, J.~H.~Godwin, G.~R.~Jehu and W.~B.~Perkins,
  Phys.\ Lett.\ B {\bf 771} (2017) 230
  doi:10.1016/j.physletb.2017.05.052
  [arXiv:1702.08273 [hep-th]].


\bibitem{Narison:1983kn}
  S.~Narison and R.~Tarrach,
  Phys.\ Lett.\  {\bf 125B} (1983) 217.
  doi:10.1016/0370-2693(83)91271-6
  
\bibitem{Morozov:1985ef}
  A.~Y.~Morozov,
  Sov.\ J.\ Nucl.\ Phys.\  {\bf 40} (1984) 505
   [Yad.\ Fiz.\  {\bf 40} (1984) 788].
  
\bibitem{Bauer:1997gs}
  C.~W.~Bauer and A.~V.~Manohar,
  Phys.\ Rev.\ D {\bf 57} (1998) 337
  doi:10.1103/PhysRevD.57.337
  [hep-ph/9708306].
  
\bibitem{Gracey:2002he}
  J.~A.~Gracey,
  Nucl.\ Phys.\ B {\bf 634} (2002) 192
   Erratum: [Nucl.\ Phys.\ B {\bf 696} (2004) 295]
  doi:10.1016/S0550-3213(02)00334-6, 10.1016/j.nuclphysb.2004.06.053
  [hep-ph/0204266].
 
 

\bibitem{PassVelt} G. Passarino and M. Veltman, Nucl. Phys. B {\bf
    160}, 151, (1979).


\bibitem{Kunszt:1994np}
  Z.~Kunszt, A.~Signer and Z.~Trocsanyi,
  Nucl.\ Phys.\ B {\bf 420} (1994) 550
  doi:10.1016/0550-3213(94)90077-9
  [hep-ph/9401294].



\bibitem{Weinberg:1965nx}
  S.~Weinberg,
  Phys.\ Rev.\  {\bf 140} (1965) B516.
  doi:10.1103/PhysRev.140.B516





\bibitem{Dunbar:1995ed}
  D.~C.~Dunbar and P.~S.~Norridge,
  Class.\ Quant.\ Grav.\  {\bf 14} (1997) 351
  doi:10.1088/0264-9381/14/2/009
  [hep-th/9512084].
  

\bibitem{Broedel:2012rc}
  J.~Broedel and L.~J.~Dixon,
  JHEP {\bf 1210} (2012) 091
  doi:10.1007/JHEP10(2012)091
  [arXiv:1208.0876 [hep-th]].



\bibitem{Cohen:2010mi}
  T.~Cohen, H.~Elvang and M.~Kiermaier,
  JHEP {\bf 1104} (2011) 053
  doi:10.1007/JHEP04(2011)053
  [arXiv:1010.0257 [hep-th]].


\bibitem{Dixon:1993xd}
  L.~J.~Dixon and Y.~Shadmi,
  Nucl.\ Phys.\ B {\bf 423} (1994) 3
   Erratum: [Nucl.\ Phys.\ B {\bf 452} (1995) 724]
  doi:10.1016/0550-3213(94)90563-0, 10.1016/0550-3213(95)00450-7
  [hep-ph/9312363].





\bibitem{He:2016iqi} 
  S.~He and Y.~Zhang,
  JHEP {\bf 1702}, 019 (2017)
  doi:10.1007/JHEP02(2017)019
  [arXiv:1608.08448 [hep-th]].


  
\bibitem{Britto:2005fq}
  R.~Britto, F.~Cachazo, B.~Feng and E.~Witten,
  Phys.\ Rev.\ Lett.\  {\bf 94} (2005) 181602
  [hep-th/0501052].


\bibitem{Norridge:1996he}
  P.~S.~Norridge,
  Phys.\ Lett.\ B {\bf 387} (1996) 701
  doi:10.1016/0370-2693(96)01109-4
  [hep-th/9606067].
  
\bibitem{Bern:1994zx}
  Z.~Bern, L.~J.~Dixon, D.~C.~Dunbar and D.~A.~Kosower,
  Nucl.\ Phys.\ B {\bf 425} (1994) 217
  doi:10.1016/0550-3213(94)90179-1
  [hep-ph/9403226].

\bibitem{Bern:1994cg}
  Z.~Bern, L.~J.~Dixon, D.~C.~Dunbar and D.~A.~Kosower,
  Nucl.\ Phys.\ B {\bf 435} (1995) 59
  doi:10.1016/0550-3213(94)00488-Z
  [hep-ph/9409265].
  

 
\bibitem{vanNieuwenhuizen:1976vb}
  P.~van Nieuwenhuizen and C.~C.~Wu,
  J.\ Math.\ Phys.\  {\bf 18} (1977) 182.
  doi:10.1063/1.523128

\bibitem{Dunbar:2017qxb}
  D.~C.~Dunbar, G.~R.~Jehu and W.~B.~Perkins,
  Phys.\ Rev.\ D {\bf 95} (2017) no.4,  046012
  doi:10.1103/PhysRevD.95.046012
  [arXiv:1701.02934 ].
  
\bibitem{Bern:2017puu}
  Z.~Bern, H.~H.~Chi, L.~Dixon and A.~Edison,
  Phys.\ Rev.\ D {\bf 95} (2017) no.4,  046013
  doi:10.1103/PhysRevD.95.046013
  [arXiv:1701.02422 ].
  
  
\bibitem{Dunbar:2002gu}
  D.~C.~Dunbar and N.~W.~P.~Turner,
  Class.\ Quant.\ Grav.\  {\bf 20} (2003) 2293
  doi:10.1088/0264-9381/20/11/323
  [hep-th/0212160].


\bibitem{Fulling:1992vm}
  S.~A.~Fulling, R.~C.~King, B.~G.~Wybourne and C.~J.~Cummins,
  Class.\ Quant.\ Grav.\  {\bf 9} (1992) 1151.
  doi:10.1088/0264-9381/9/5/003

 
\bibitem{BelRob}
L.\ Bel, Acad. Sci. Paris, Comptes Rend. {\bf 247}:1094 (1958),
{\bf 248}:1297 (1959) \\
I. Robinson, unpublished 
King's College Lectures (1958), Class.\ and Quant.\ Grav.\ 
{\bf 14}:4331 (1997) 

\bibitem{R4sugra}
S.~Deser, J.H.~Kay and K.S.~Stelle,
Phys. Rev. Lett. {\bf 38}:527 (1977); \\
R.E.~Kallosh,
Phys. Lett. {\bf B99}:122 (1981); \\
P.S.~Howe, K.S.~Stelle and P.K.~Townsend,
Nucl. Phys. {\bf B191}:445 (1981).


\bibitem{DS}
S. Deser and D.\ Seminara,
Phys.\ Rev.\ Lett.\ {\bf 82}:2435 (1999) [hep-th/9812136],
Phys.\ Rev.\  {\bf D62}:084010 (2000) 
[hep-th/0002241]



\bibitem{TurnerThesis}
 N.~W.~P.~Turner, ``Ultra-Violet Infinities and Counterterms in Higher Dimensional Yang-Mills and Gravity'', Ph.D. Thesis, University of Wales, 2003. 



\bibitem{Benincasa:2007qj}
  P.~Benincasa, C.~Boucher-Veronneau and F.~Cachazo,
  JHEP {\bf 0711} (2007) 057
  doi:10.1088/1126-6708/2007/11/057
  [hep-th/0702032].


\bibitem{Risager:2005vk}
  K.~Risager,
  JHEP {\bf 0512} (2005) 003
  doi:10.1088/1126-6708/2005/12/003
  [hep-th/0508206].

\bibitem{BjerrumBohr:2005jr}
  N.~E.~J.~Bjerrum-Bohr, D.~C.~Dunbar, H.~Ita, W.~B.~Perkins and K.~Risager,
  JHEP {\bf 0601} (2006) 009
  doi:10.1088/1126-6708/2006/01/009
  [hep-th/0509016].

\bibitem{Dunbar:1999nj}
  D.~C.~Dunbar, B.~Julia, D.~Seminara and M.~Trigiante,
  JHEP {\bf 0001} (2000) 046
  doi:10.1088/1126-6708/2000/01/046
  [hep-th/9911158].

\bibitem{Green:1987mn}
  M.~B.~Green, J.~H.~Schwarz and E.~Witten,
  Cambridge, Uk: Univ. Pr. ( 1987) 596 P. ( Cambridge Monographs On Mathematical Physics)


\end{thebibliography}
\end{document}